\documentclass[aps,pre,reprint,groupedaddress]{revtex4-2}

\usepackage{graphicx}
\usepackage{bm}
\usepackage{xcolor}

\begin{document}

\title{Invading activity fronts or near-critical clusters may stabilize vulnerable excitable 
 	systems against stochastic trajectories terminating in an absorbing state}

\author{Kenneth A.~V.~Distefano} 	\email{dkenneth@vt.edu}
\author{Sara~Shabani} 				\email{sarashabani@vt.edu}
\author{Uwe~C.~T\"auber\,} 			\email{tauber@vt.edu}  
\altaffiliation{Faculty of Health Sciences, Virginia Tech}
\affiliation{Department of Physics (MC 0435) \& Center for Soft Matter and Biological Physics, 
    		 Robeson Hall, 850 West Campus Drive, Virginia Tech, Blacksburg, VA 24061}

\date{\today}

\begin{abstract}
Stochastic chemical reaction or population dynamics in finite systems often terminates in an 
absorbing state. 
Yet in large spatially extended systems, the time to reach the absorbing configuration 
({\em e.g.}, species extinction or fixation) becomes exceedingly long. 
Tuning control parameters may effectively diminish the survival probability, rendering active 
states (species coexistence) susceptible to stochastic extinction events. 
In inhomogeneous settings, where a vulnerable subsystem is diffusively coupled to an 
adjacent stable patch, the former is re-activated through influx from the interfaces, provided 
the vulnerable region either sustains spreading activity fronts or near-critical large active 
clusters. 
Either dynamical structure resets the system to correlated initial conditions favorable for
survival.  
We demonstrate this generic elimination of finite-size extinction instabilities via immigration 
flux in stochastic predator-prey, cyclic competition, and epidemic spreading models.
\end{abstract}

\maketitle

% Introduction: absorbing states.
Stochastic dynamics captures the essence of many processes in nature, at least on a 
coarse-grained description level.
In genuine non-equilibrium physical, chemical, and biological systems, {\em absorbing 
states} play a prominent role \cite{Kampen81}.
Once such a configuration is reached, the transition rate to any other state vanishes.
Consequently, the stochastic kinetics ceases and escape from an absorbing state is 
impossible.
Prominent examples are inert compounds in chemical kinetics \cite{Lindenberg20}; 
population extinction or fixation in ecology \cite{May73, Hofbauer98}; pathogen eradication 
or full immunity in epidemiology \cite{Keeling11}; but also trapped configurations in 
constrained or disordered physical systems \cite{Jack06, Kohl16, Deger22}; and certain 
dynamical phases in fluid turbulence \cite{Takeuchi07, Shih16}. 

% Main message.
In this Letter, we consider a spatially extended system that is prone to reach its absorbing 
state through stochastic fluctuations.
We demonstrate how the ensuing extinction or fixation events can be avoided through 
diffusive coupling with another more stable system via either of two related but distinct
invasive mechanisms:
(a) emerging activity fronts that propagate through the vulnerable patch and revitalize it;
(b) near-critical active invasion clusters that permeate the endangered region.
In both cases, the initial conditions in the extinction-prone subsystem are reset to spatially
correlated configurations that efficiently promote resilience of the active state.

% Background: stochastic extinction in finite systems.
We begin by explaining the relevant background.
First, we recall that any {\em finite} stochastic system featuring one or more absorbing 
states will ultimately terminate there \cite{Kampen81}.
Chemical activity, species coexistence, and epidemic outbreaks are thus fundamentally
transient phenomena.
However, mean times $t_\mathrm{ext}$ to reach extinction or fixation typically grow 
exponentially with system size, {\em i.e.}, the number of constituents $N$: 
$t_\mathrm{ext}(N) \sim e^{c N}$ \cite{Reichenbach07, Parker09, He11, Dobrinevsky12}.
In practice, therefore, stochastic trajectories in large enough systems that are initiated at 
sufficient separation from an absorbing state become highly unlikely to ever reach it.
Moreover, in {\em spatially extended} systems \cite{Durrett99} spreading activity fronts tend 
to locally trigger excitations and facilitate escape from absorbing states \cite{Reichenbach07,
Moreno25}.
Thus, mean extinction times are drastically prolonged, and in effect, sufficiently big systems 
display persistent activity for any experimentally or computationally accessible time scales 
$t < t_\mathrm{ext}$.
Indeed, many prominent (mean-field) reaction-diffusion models sustain invasion fronts that 
propagate from either localized or extended active sources into dormant inactive domains
\cite{Murray02, Cross09}.
In stochastic reactive systems, demographic fluctuations may excite robust dynamical 
{\em noise-induced spatio-temporal patterns} \cite{Mobilia07, Butler09, Tauber14,
Dobramysl18, Tauber24}.

% Genuine active-to-absorbing state phase transitions.
Taking first the {\em infinite-system} limit $N \to \infty$ followed by $t \to \infty$, one may
encounter the scenario that upon changing a relevant control parameter ({\em e.g.}, a
reaction rate), a genuine {\em dynamical phase transition} occurs that separates an active 
stationary phase from an inactive, absorbing configuration. 
Since the latter manifestly violates micro-reversibility and detailed balance, such
active-to-absorbing state transitions are situated away from thermal equilibrium.
On regular spatial lattices, they are generically continuous and universally governed by the 
scaling exponents of critical {\em directed percolation} (DP) \cite{Janssen81, Grassberger82, 
Hinrichsen00, Odor04, Henkel08, Tauber14, Tauber17}, even in multi-species systems 
\cite{Janssen97}.
Typical examples are the contact process \cite{Harris74, Marro99} or stochastic 
susceptible-infectious-susceptible (SIS) epidemic model on a lattice \cite{Keeling11} and the 
stochastic spatial Lotka--Volterra (LV) predator-prey or pathogen-host competition model 
\cite{Matsuda92, Satulovsky94} with finite prey carrying capacity $K$ that induces a predator extinction threshold \cite{Lipowski00, Monetti00, Antal01, Mobilia07, Dobramysl18}.
Exceptions involve higher-order processes beyond binary reactions, the presence of 
additional conserved (diffusive) modes, and/or persistent memory effects that imply 
temporal non-locality \cite{Hinrichsen00, Odor04, Henkel08, Tauber14}. 
For example, the epidemic threshold in the susceptible-infectious-recovered (SIR) model 
\cite{Kermack27, Murray02, Keeling11} is characterized by the {\em dynamic isotropic 
percolation} (DIP) universality class \cite{Grassberger83, Cardy85, Tome10}.

\begin{figure}
\includegraphics[width=0.45\columnwidth]{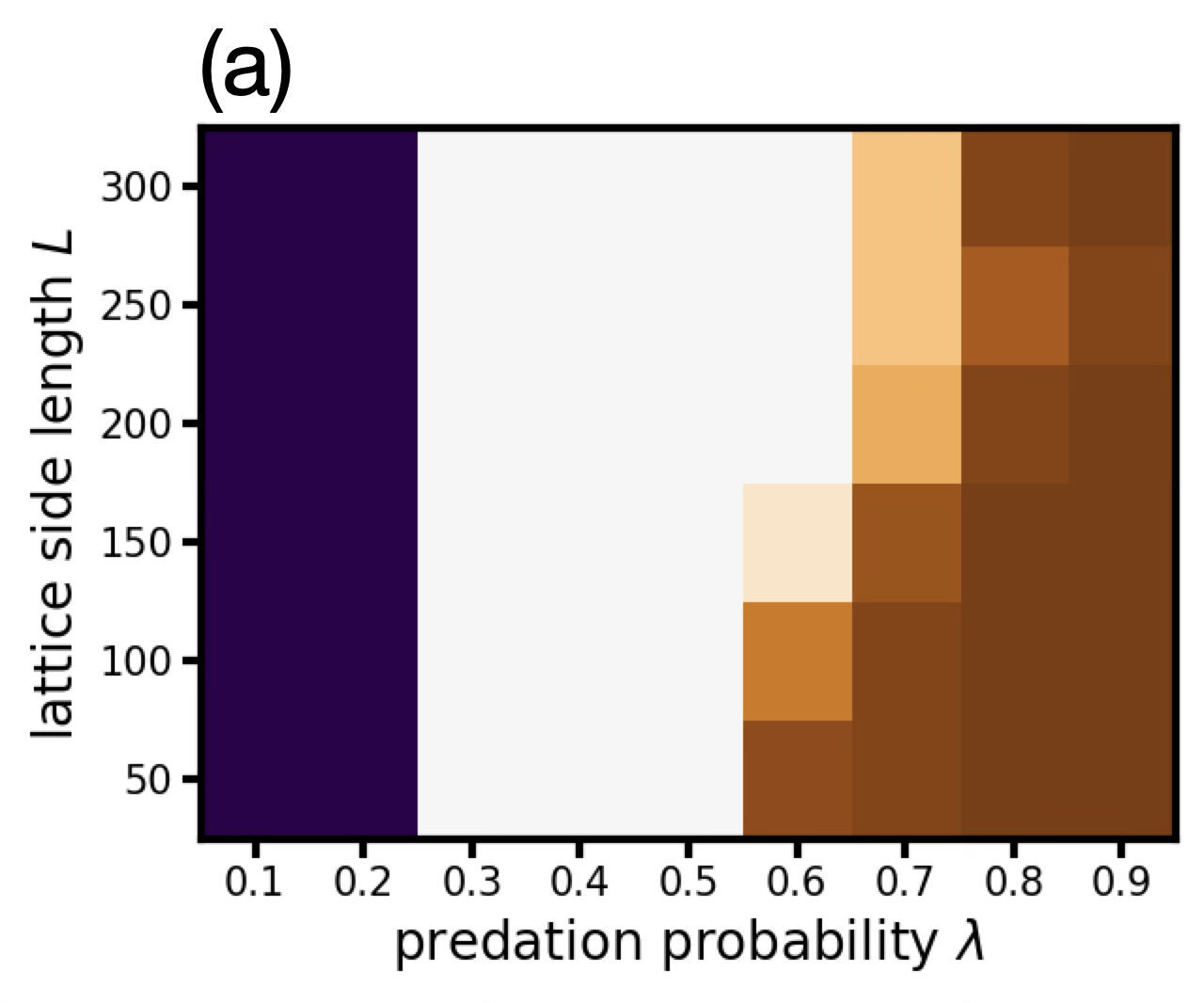}
\includegraphics[width=0.52\columnwidth]{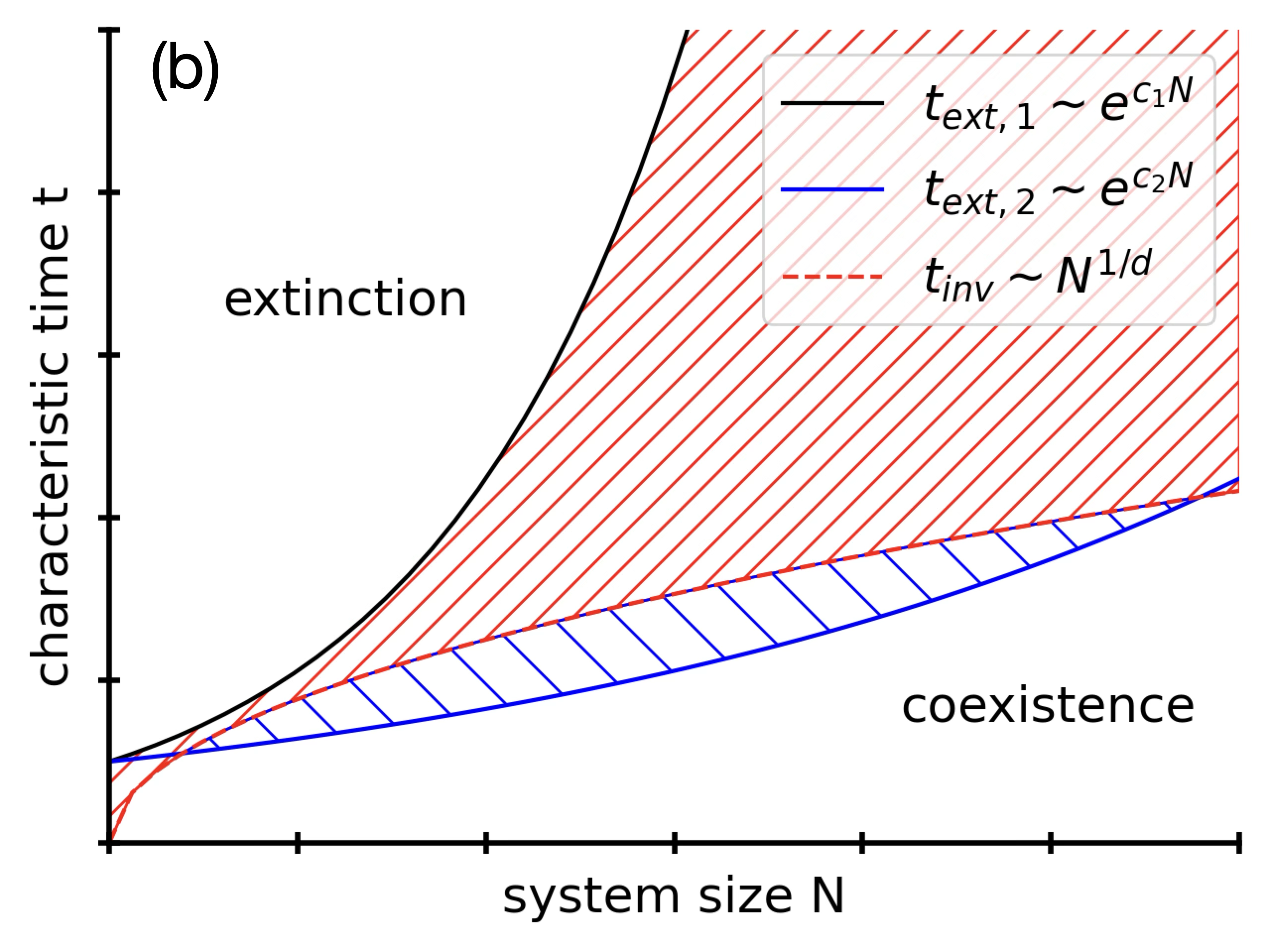}
\vskip -0.3cm
\caption{(a) Coexistence (white), prey fixation (dark blue) and total extinction (with increasing
	probability indicated in darker brown shades) regimes in the stochastic LV model on a 
	$L \times L$ square lattice as function of $L$ and the predation probability $\lambda$ 
	(at fixed $\mu = 0.1$, $\sigma = 0.2$, and $K = 3$).
	The (largely) size-independent genuine DP predator extinction (prey fixation) transition
	of type (i) occurs at $\lambda_c \approx 0.25$.
	For large $\lambda > 0.5$, one observes stochastic type-(ii) total population extinction 
	events with strongly $L$-dependent probability (brown) that cannot be captured in 
	mean-field theory \cite{Satulovsky94,Swailem23, Distefano26}. 
	(b) Schematic dependence on system size $N$ of the invasion front traversal time 
	$t_\mathrm{inv} \sim N^{1/d}$ (red dashed) and mean extinction times 
	$t_\mathrm{ext}$ for robust (black, $c_1 > c_2$) and vulnerable (sub-)systems (blue) 
	prone to be stochastically driven to an absorbing state.
	For $t_\mathrm{ext,1}> t > t_\mathrm{inv}$ (shaded red area), an excitable patch that
	in isolation is vulnerable to stochastic extinction for $t > t_\mathrm{ext,2}$ (shaded 
	blue) may be re-animated and stabilized via diffusive coupling to an active region. 
\label{fig:extsch}} \vskip -0.5cm
\end{figure}

% General picture.
Thus, in non-linear stochastic systems, one may observe two {\em distinct} types of 
{\em extinction} (or fixation) {\em ``transitions"}:
(i) {\em Genuine active-to-absorbing state phase transitions} that require taking {\em first}
the thermodynamic limit $N \to \infty$, then $t \to \infty$, and tuning appropriate control 
parameters.
They display only a weak dependence of the transition point on $N$ through standard 
finite-size scaling.
(ii) {\em Stochastic extinction events in finite (``small'') systems}, triggered by temporal 
trajectories that randomly reach an absorbing state.
Aside from adjusting control parameters, these are highly sensitive to the system size $N$ 
and potentially to the chosen initial conditions. 
Upon increasing $N$, the extinction probability rapidly decreases and vanishes in the 
thermodynamic limit, where a persistent active phase prevails.
Consequently, type-(ii) extinction or fixation events technically are ``finite-size" phenomena: 
They are limited to comparatively small systems, crucially rely on demographic internal 
fluctuations, and cannot be captured by any mean-field approach.
Yet they must surely feature prominently in real biological, ecological, chemical, and 
physical systems.
For the paradigmatic LV model, this is demonstrated in Fig.~\ref{fig:extsch}(a), which shows 
both the ``true'' type (i) prey fixation transition {\em and} a strongly size-dependent type (ii) 
total population extinction at large predation efficacies $\lambda$ in a square lattice of linear 
size $L$ \cite{Satulovsky94, Swailem23, Distefano26}. 

% Spatially inhomogeneous systems.
We now turn to {\em spatially inhomogeneous systems} consisting of diffusively coupled 
regions set in distinct macroscopic states. 
If these are either active or absorbing configurations separated by a type-(i) genuine phase 
transition, they remain essentially decoupled: 
Their mutual influence is limited to the vicinity of the interfaces between the subsystems, on 
the scale of the respective characteristic correlation lengths.
This is manifested, for example, in checkerboard LV model patches alternatingly situated in 
the two-species coexistence phase and the predator extinction regime \cite{Heiba18}, and in 
similarly coupled subsystems comprising two different cyclic three-species competition 
models \cite{May73, Hofbauer98}, namely the ``rock-paper-scissors'' and May--Leonard (ML) 
\cite{May75} variants \cite{Arnau20}.

In contrast, here we show that diffusively coupling two finite regions that are respectively 
vulnerable to / robust against type-(ii) stochastic extinction or fixation events, induces activity 
influx across the system interfaces from the stable patch. 
If the inactive region sustains traveling invasion fronts or large active clusters, these 
subsequently re-invigorate the {\em entire} unstable subsystem and persistently render it 
immune to reaching the absorbing state.
This is true even if the incitatory patch is much smaller in its spatial extent than the 
vulnerable region, and after the diffusive coupling of both subsystems is terminated.
Stochastic extinction events are countermanded by another finite-size effect, namely 
population influx through the interfaces.
The robust patch merely needs to remain in a stable active state itself and stay in contact 
with the vulnerable subsystem until the invasion fronts have permeated enough to re-incite 
activity. 
The in isolation extinction-prone region is then characterized by the same macroscopic 
spatio-temporal features that it would display if its extension were increased sufficiently to 
move it towards large extinction times $t_\mathrm{ext} \gg t$ \cite{Serrao21, Distefano26,
Shabani26}.

% Context: source-sink kinetics and refuges, mean-field invasion fronts.
Of course, the relevance of refuges for the survival of vulnerable populations and the 
associated source-sink dynamics has long been recognized in biology and ecology 
\cite{McNair84, Dias96, Furrer16}.
In the standard mean-field reaction-diffusion equation modeling of such situations, robust 
invasion fronts occur quite generically \cite{Murray02}.
However, we emphasize that this is due to the extinct, absorbing state invariably being 
unstable in the rate equation framework, at stark variance with a proper stochastic treatment
that adequately captures internal reaction noise as well as the discrete nature of interacting /
reacting particle systems.

% {\em Relevant time scales.}
To illustrate the qualitative picture in our {\em stochastic} spatially inhomogeneous system, 
Fig.~\ref{fig:extsch}(b) schematically depicts relevant time scales:
Stabilization of the vulnerable region (2) is attainable for simulation (or experiment) run 
times $t < t_\mathrm{ext,1} \sim e^{c_1 N_1}$, the stable patch's (1) mean extinction
time (black curve).
Beyond the weaker subsystem's extinction time $t > t_\mathrm{ext,2} \sim e^{c_2 N_2}$ 
($0 < c_2 < c_1$, blue line), activity in the vulnerable region will have ceased.
Traveling immigration waves (with ``pulled'' front speed $v$ \cite{Distefano26}) originating 
at both interfaces with the stable patch will traverse through the inactive region of 
perpendicular width $L_\perp$ by the characteristic time 
$t_\mathrm{inv} \approx L_\perp / 2 v \sim N_2^{1/d}$ (red-dashed).
During the ``transient" window $t_\mathrm{inv} < t < t_\mathrm{ext,1}$, influx from the 
robust region 1 countermands stochastic extinction in the vulnerable region 2, irrespective of 
its width $L_\perp$.

% Hierarchical population dynamics: Lotka-Volterra predator-prey model.
We exemplify this general stabilization mechanism in the simplest hierarchical population 
dynamics consisting of just two competing species, namely the LV model defined through the 
stochastic processes predator (pathogen) death $I \to \emptyset$, prey (host) reproduction 
(branching) $S \to S + S$, and the predation (infection) reaction $S + I \to I + I$, 
respectively with probabilities (or positive rates, per unit time) $\mu$, $\sigma$, and 
$\lambda$.
On a regular square lattice with periodic boundary conditions, we impose a finite local 
carrying capacity restricting all site-$i$ occupation numbers to 
$n_i = n_{S\,i} + n_{I\,i} \leq K$.
The algorithm employed in this work \cite{Distefano26} places prey offspring adjacent to 
their parent's location $i$ provided $n_j \leq K - 1$ there, thus inducing diffusive 
propagation. 
The binary predation reaction requires both $S$ and $I$ individuals to be simultaneous 
present either on the same or a nearest-neighbor site.
The simulation advances through random sequential updates.
A Monte Carlo step (MCS) is completed when on average every particle had the chance to
undergo a reaction \cite{Mobilia07, Dobramysl18, Tauber24}.
We set $\mu = 0.1$, $\sigma = 0.2$, and $\lambda = 0.9$; for carrying capacity 
$K = 1$, systems of linear size $L = 100$ or larger then display species coexistence 
throughout the simulation runs, irrespective of their initialization.
In contrast, for $K = 3$ and a random (Poisson) initial particle placement at sufficiently 
large starting density, early-time large-amplitude local population spikes tend to drive an 
isolated system stochastically to one of its absorbing states (typically total extinction rather 
than prey fixation) even at $L = 200$ \cite{Swailem23, Distefano26}.

\begin{figure}
\includegraphics[width=0.43\columnwidth]{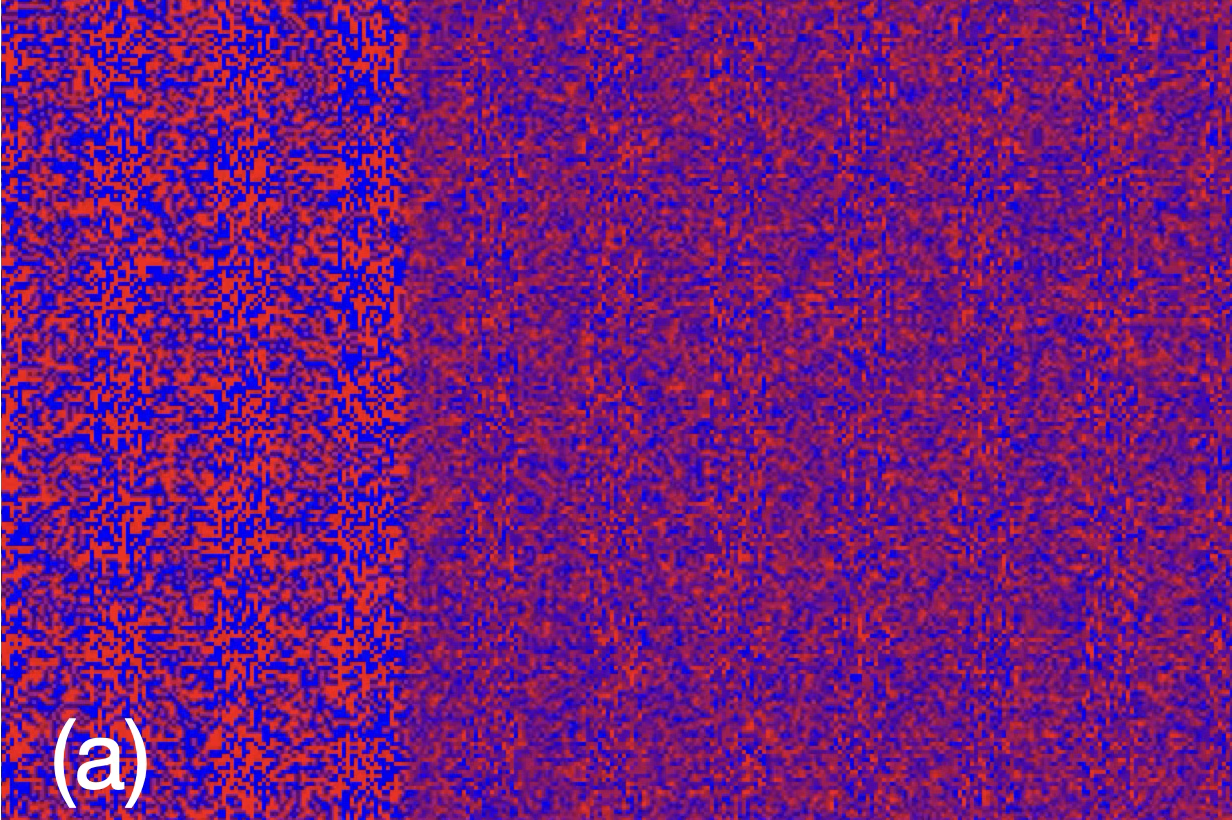} \,
\includegraphics[width=0.43\columnwidth]{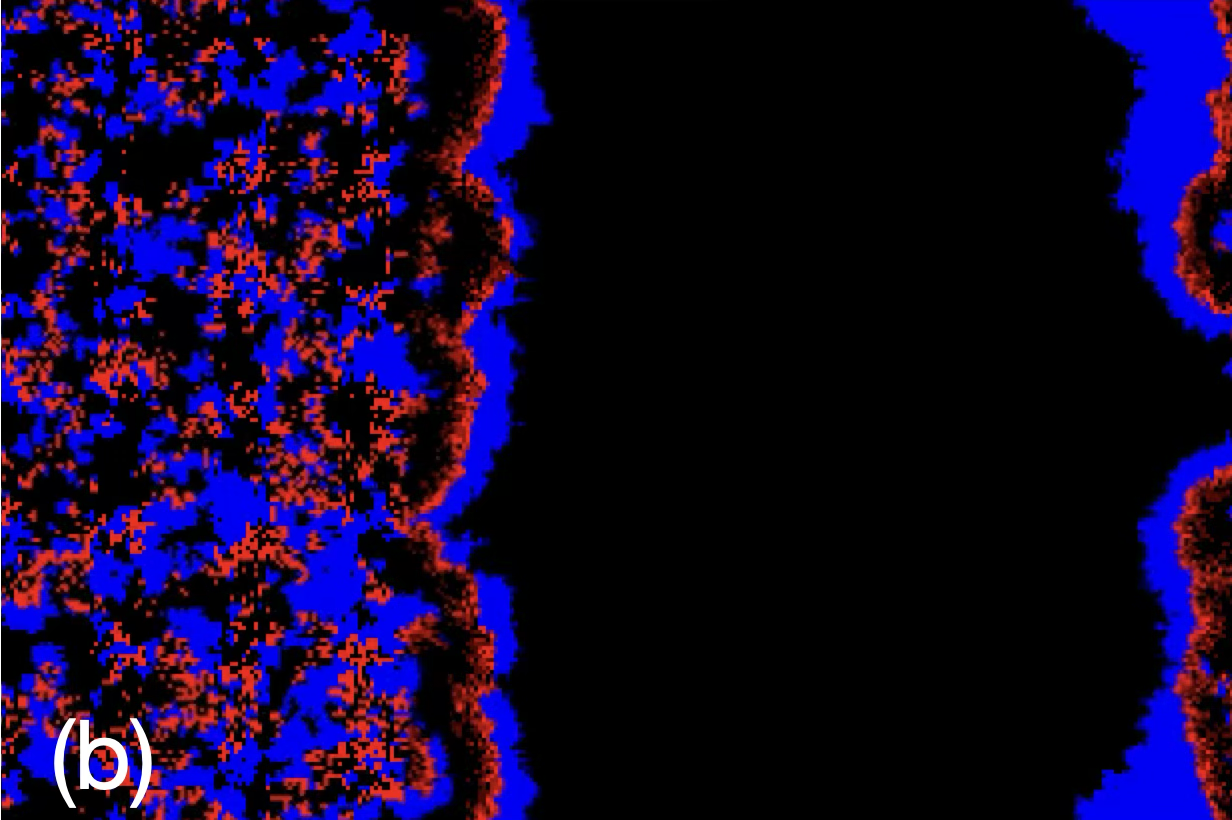} \\ \smallskip
\includegraphics[width=0.43\columnwidth]{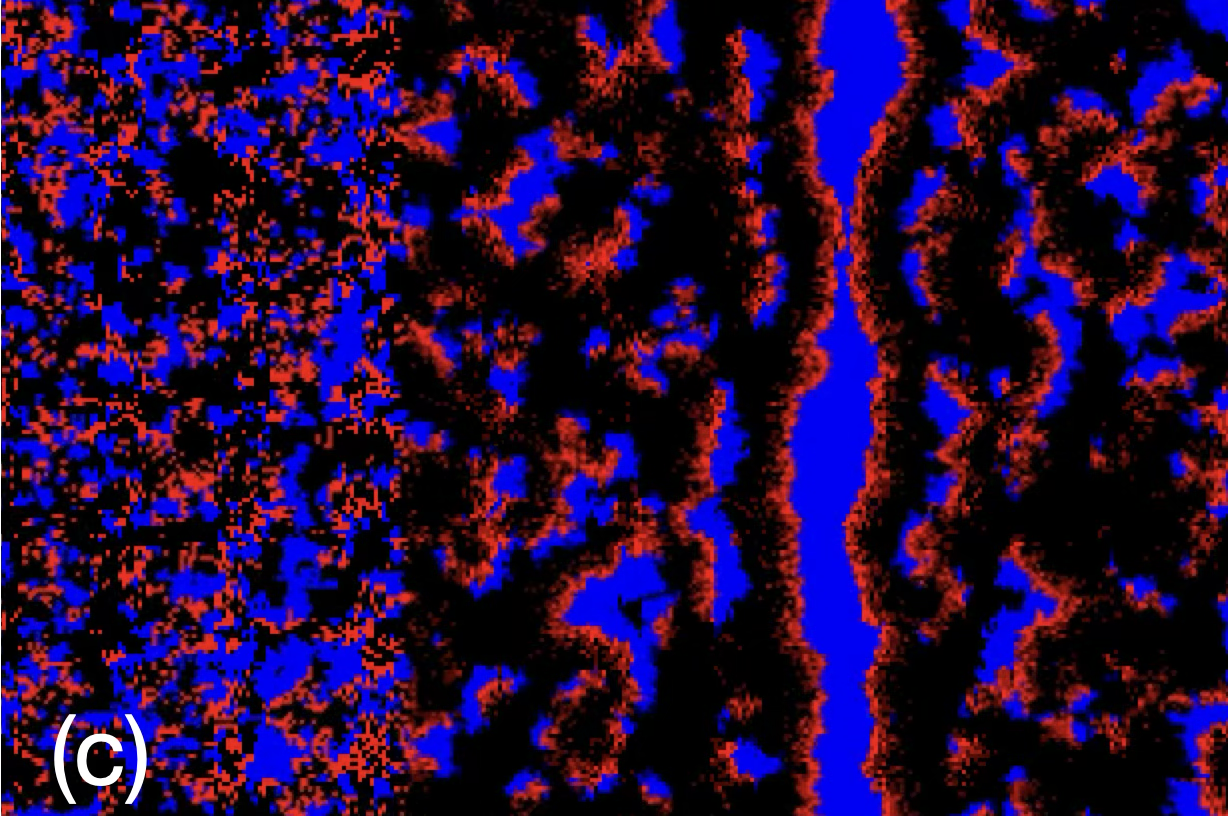} \,
\includegraphics[width=0.43\columnwidth]{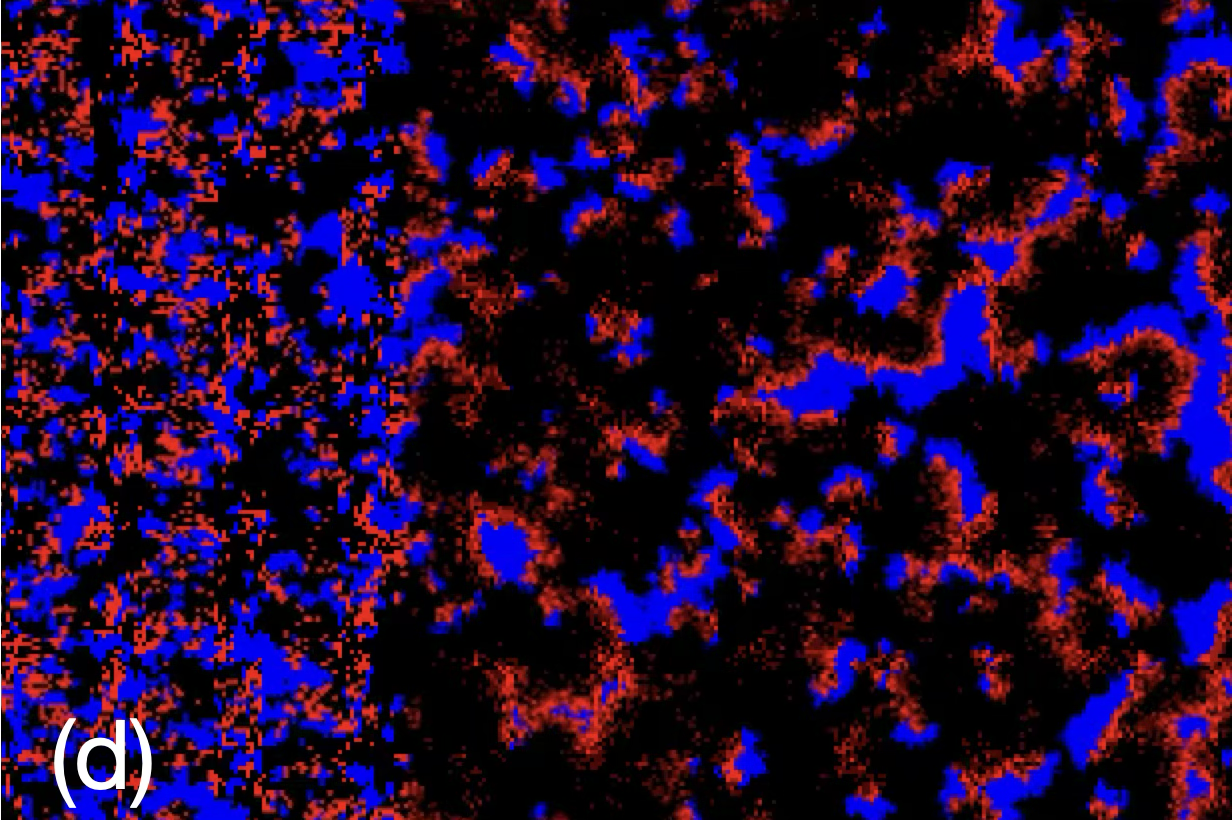} \\ 
\includegraphics[width=0.47\columnwidth]{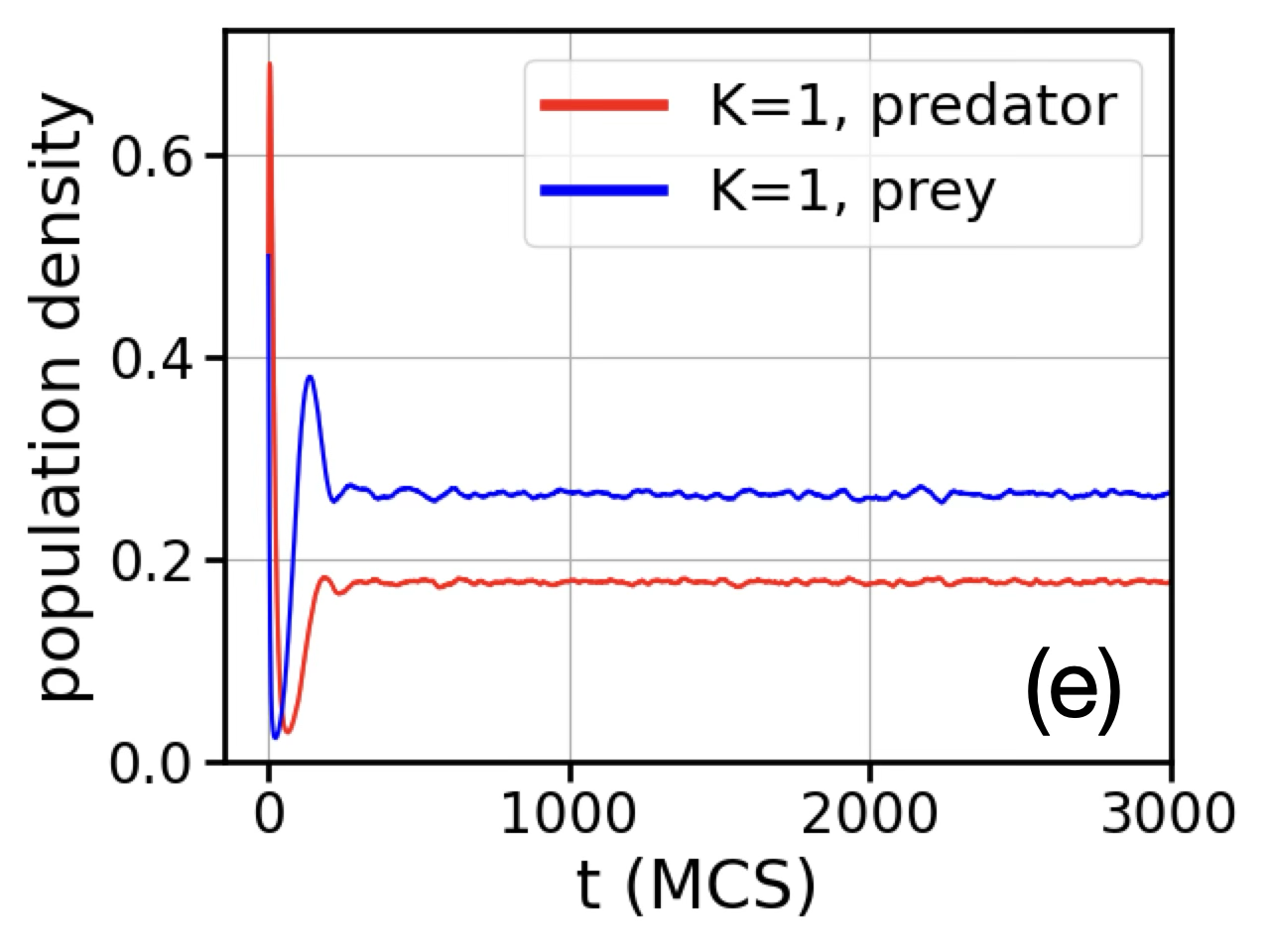}
\includegraphics[width=0.47\columnwidth]{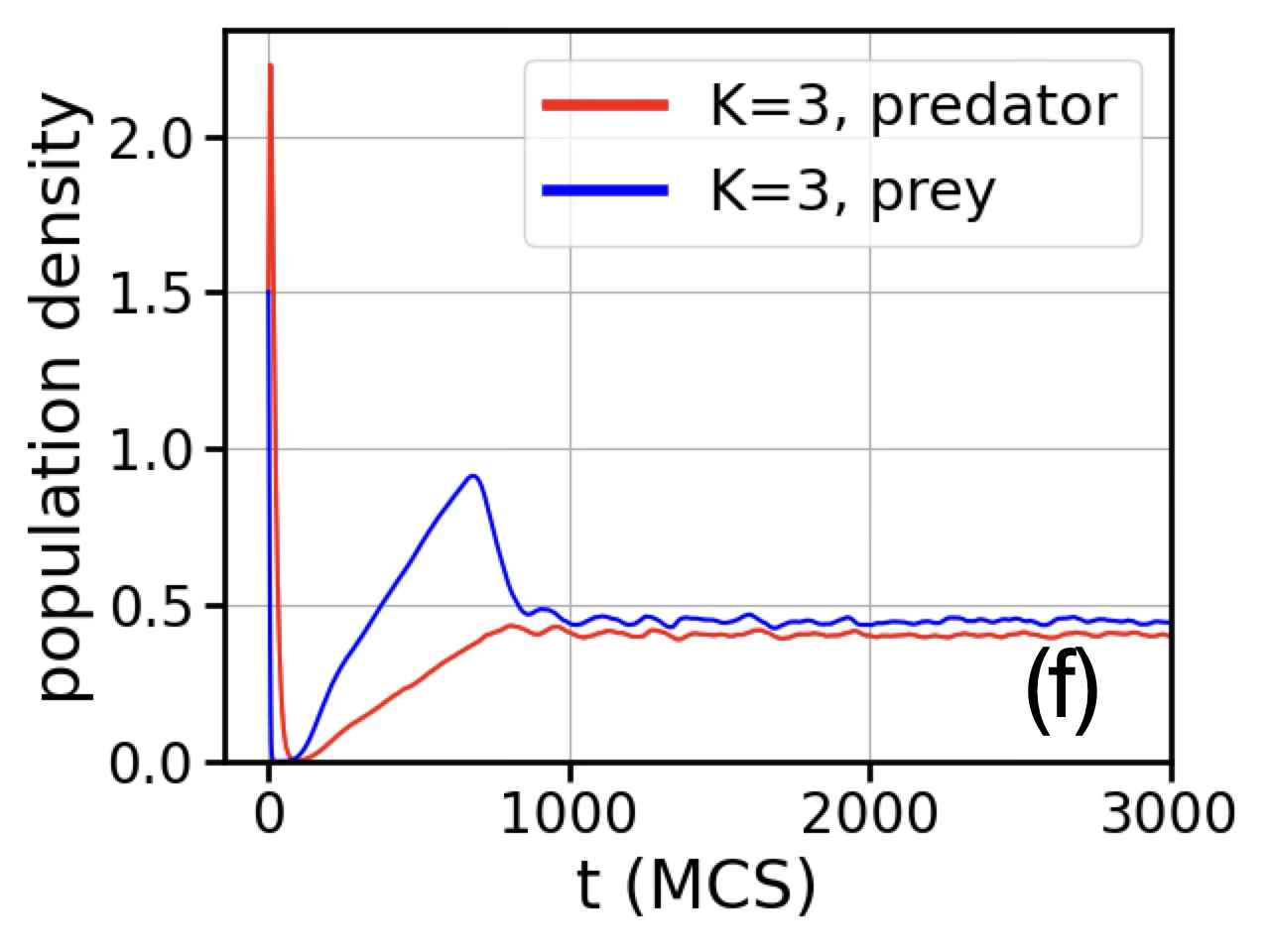}
\vskip -0.3cm
\caption{Snapshots of a single Monte Carlo simulation run for two diffusively coupled LV
	systems on a square lattice (height $L_\parallel = 200$, periodic boundary conditions),
	(a) initialized ($t = 0$) with a random distribution of single predator (red) and prey
	(blue) individuals.
	The smaller (left) patch (width $L_\perp = 100$, carrying capacity $K = 1$) resides in 
	a stable active predator-prey coexistence state characterized by persistent radially
	propagating activity waves.
	The larger (right) subsystem ($L_\perp = 200$, $K = 3$) is prone to fluctuation-driven 
	total population extinction driven by initially abundant local predation events; as has 
	happened by (b) $t = 260$ MCS (empty sites: black).
	Yet roughly planar invasion fronts emanating from both interfaces with the active 
	region enter the empty, excitable subsystem, collide at (c) $t = 800$ MCS, generate 
	wider population waves, and subsequently stabilize the vulnerable area in perpetuity 
	at the higher species densities pertinent to $K = 3$, as shown at (d) $t = 3000$ MCS)
	\cite{Distefano26, Movies}.
	Time evolution of the population densities averaged over $21$ independent runs in 
	the (e) stable and (f) vulnerable regions. 
\label{fig:LVsnap}} \vskip -0.8cm
\end{figure}
With the above fixed reaction parameters, we construct a {\em spatially inhomogeneous} 
setup where a stable LV two-species coexistence environment with $K = 1$ is placed in 
diffusive contact with a vulnerable ecology with $K = 3$:
Once a particle attempts to cross either of the two interfaces to the other region, it is 
subject to the carrying capacity there.
The simulation snapshots in Fig.~\ref{fig:LVsnap}(a)-(d) depict the ensuing dynamics in a 
typical realization:
Beginning with a random particle distribution, the smaller ($200 \times 100$) stable region 
quickly develops the characteristic LV activity fronts (left patch), whereas the population 
becomes extinct in the larger ($200 \times 200$) subsystem with $K = 3$ (right).
Yet at $t = 260$ MCS, one observes prey-predator fronts emanating from the stable 
environment into the empty patch, and as they propagate ballistically with speed $v$, 
revitalization occurs \cite{Distefano26}.
Indeed, the effect of the invasion waves may be viewed as a reset to {\em correlated}
starting conditions that avoid large destabilizing population spikes.
At $t = 800$ MCS, the two fronts collide and subsequently generate the dynamical 
quasi-stationary state characteristic for a (larger) stable LV system with $K = 3$, \textit{viz.}
with higher population densities and wider activity waves than for $K = 1$.
Graphs (e), (f) display the areal population densities $n_{A/B}(t)$ in the stable ($K = 1$) 
and vulnerable ($K = 3$) regions.
Remarkably, this {\em stabilization through invasion waves} into the inactive excitable 
domain remains effective essentially independent of its size $L_\perp$ provided it 
exceeds the typical wave front width $\xi$, and even stays intact when both regions are 
decoupled after elapsed time $t_\mathrm{dec}$, as long as a sufficient fraction of the 
vulnerable area is reactivated; i.e., $t_\mathrm{dec}$ is of the order of the traversal time 
$t_\mathrm{inv}$ \cite{Distefano26, Movies}.

% Cyclic three-species competition: asymmetric May-Leonard model.
The above stabilization mechanism of finite spatially extended systems prone to be driven
stochastically to an absorbing state is not restricted to hierarchical binary interaction schemes.
For example, consider the three-species ML cyclic competition model \cite{May75}, defined 
by the stochastic predation / reproduction reactions $S_i + S_{i+1} \to S_i$ (probability 
$\lambda_i$), $S_i \to S_i + S_i$ (probability $\sigma_i$), with species indices $i = 1,2,3$ 
and $i+3$ to be identified with $i$.
For sufficiently large system size, the ML model displays spontaneously emerging spiral 
structures that stabilize species coexistence \cite{Reichenbach07}.
Yet when the typical spiral extension reaches the system boundaries, this spatial 
``advantage" disappears, the model effectively becomes zero-dimensional and susceptible to 
stochastic fixation events, leaving just a single species \cite{He11}.
This vulnerability against two-species extinction is enhanced in strongly asymmetric ML 
models (e.g., setting $\lambda_1 \gg \lambda_2 = \lambda_3$ or 
$\lambda_1 \ll \lambda_2 = \lambda_3$), for which the stochastic trajectories approach the 
absorbing edges in parameter space (akin to heteroclinic cycles in mean-field theory).
As demonstrated in Ref.~\cite{Serrao21}, diffusively coupling an asymmetric cyclic ML region
with an even much smaller stable patch ({\em viz.}, with symmetric reactivities) incites planar 
invasion fronts at the subsystems' interfaces that in turn generate spiral structures in the bulk 
of the vulnerable region and thereby re-animate all three competing populations. 

% Epidemic spreading: susceptible-infectious-susceptible and -recovered models.
Finally, we turn to basic epidemic processes \cite{Tauber14} captured by the stochastic 
spatial SIS system \cite{Murray02, Keeling11}, where the infection spreads via the same 
predation reaction $S + I \to I + I$ (probability $\lambda$) as in the LV model, but with the 
birth/death processes replaced by the single recovery reaction $I \to S$, to which we 
assign a probability $\gamma$.
Consequently, the {\em total} particle number and area density $n_S + n_I$ remain 
conserved under this stochastic dynamics and full population extinction is prohibited.
Similar to the LV model, the SIS system features an active-to-absorbing transition pertaining 
to the $I$ species in the DP universality class. 
The active phase permits repeated infection fronts spreading through the system.
Here we fix $\gamma = 0.1$, restrict infection processes to neighboring lattice sites, and 
tune the infection probability close to the (infinite-system) epidemic threshold 
$\lambda_c \approx 0.108$.
Thus for $\lambda = 0.115$, even in small systems with linear extension $L = 100$,
the majority of simulation runs reaches the $S$-$I$ coexistence state. 
Yet among $100$ realizations initialized with a single (or a few) infectious $I$ seed(s) 
surrounded by abundant susceptible individuals $S$, we found $27$ runs to terminate at 
disease extinction, $n_I \to 0$.
In contrast, for a slightly reduced infectivity $\lambda = 0.109$, {\em all} simulation runs 
are stochastically driven to the absorbing state where the infection wave dies out quickly
\cite{Shabani26}.

\begin{figure}
\includegraphics[width=0.47\columnwidth]{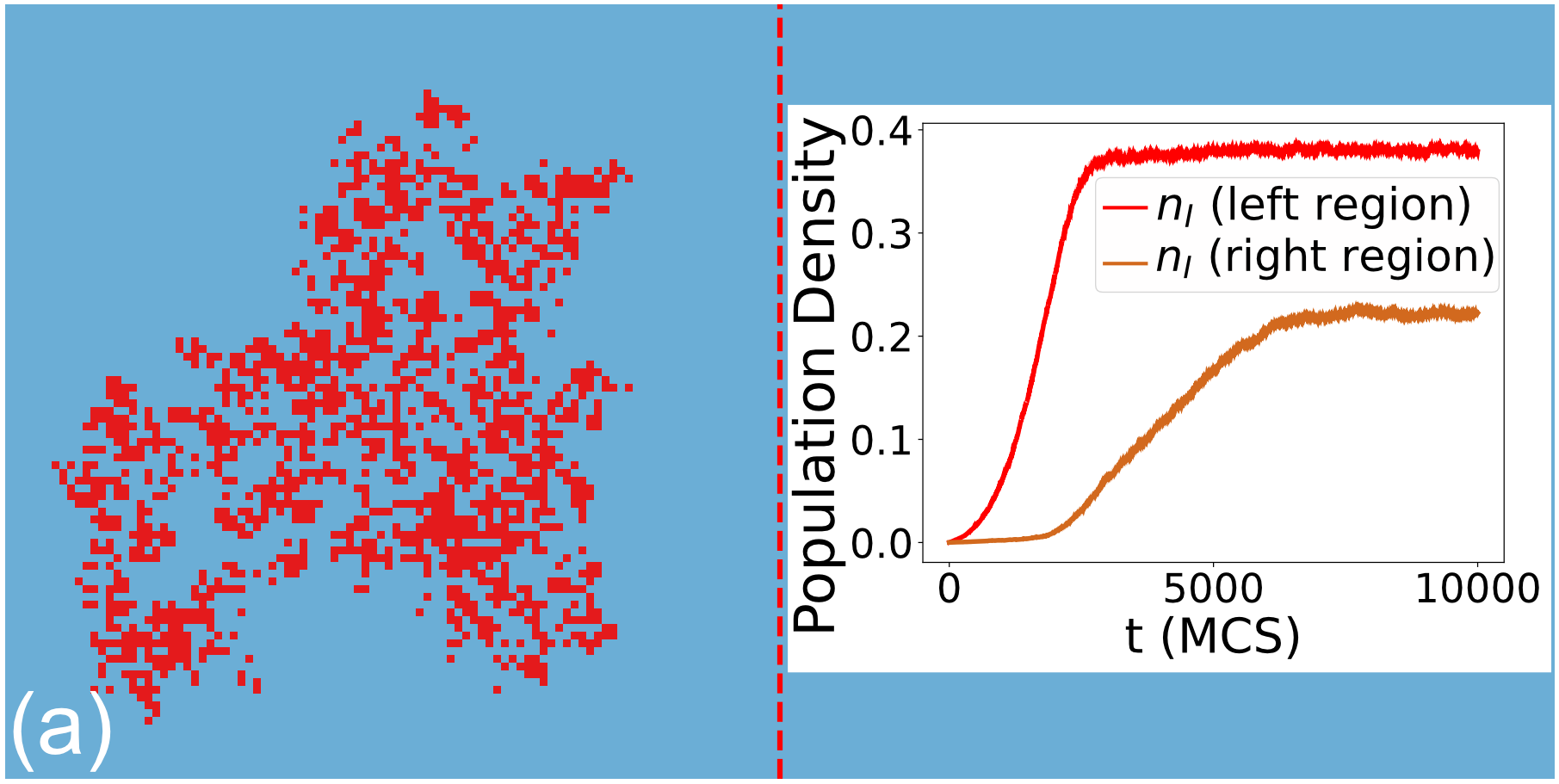} \,
\includegraphics[width=0.47\columnwidth]{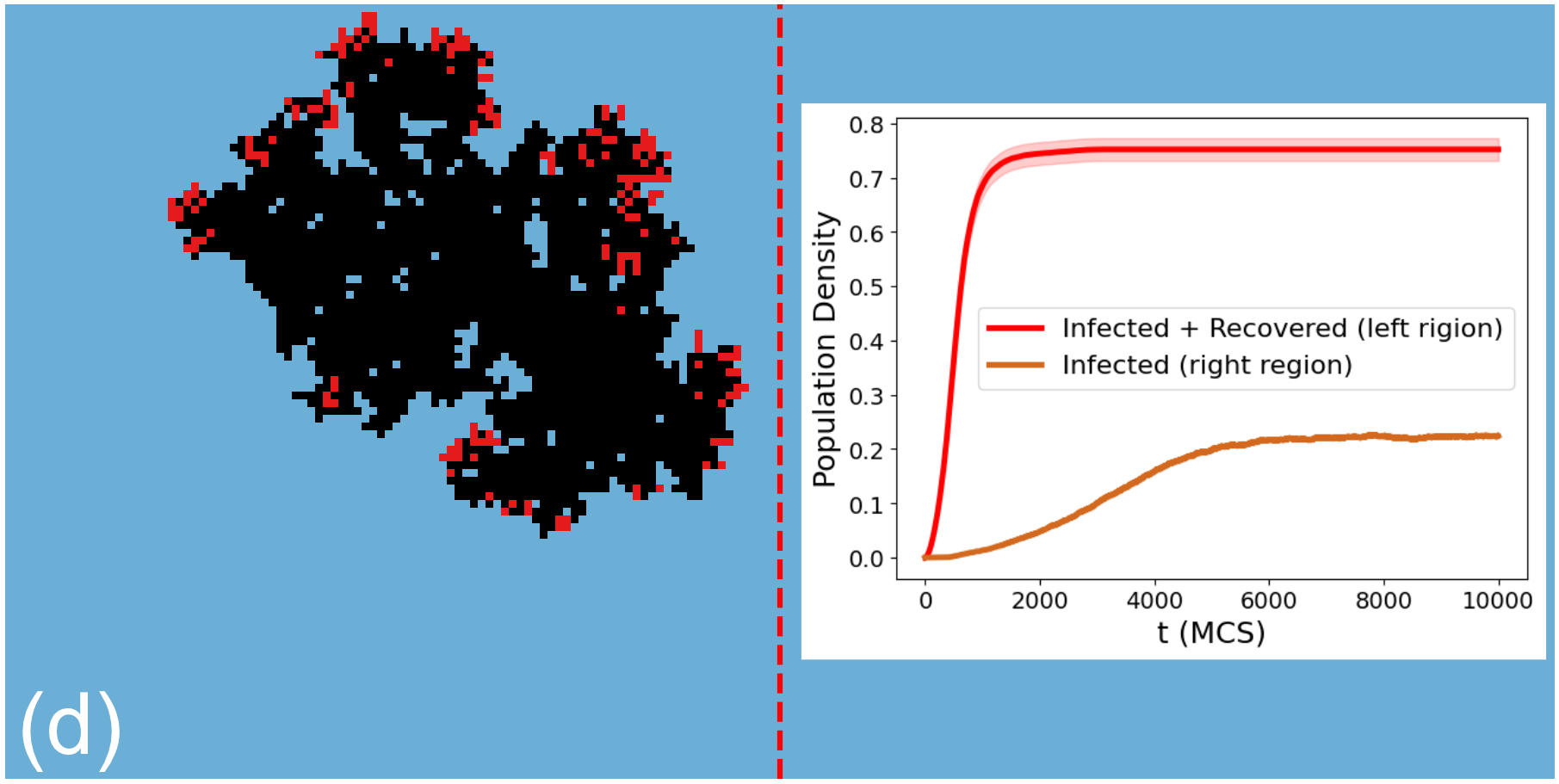} \\ \smallskip
\includegraphics[width=0.47\columnwidth]{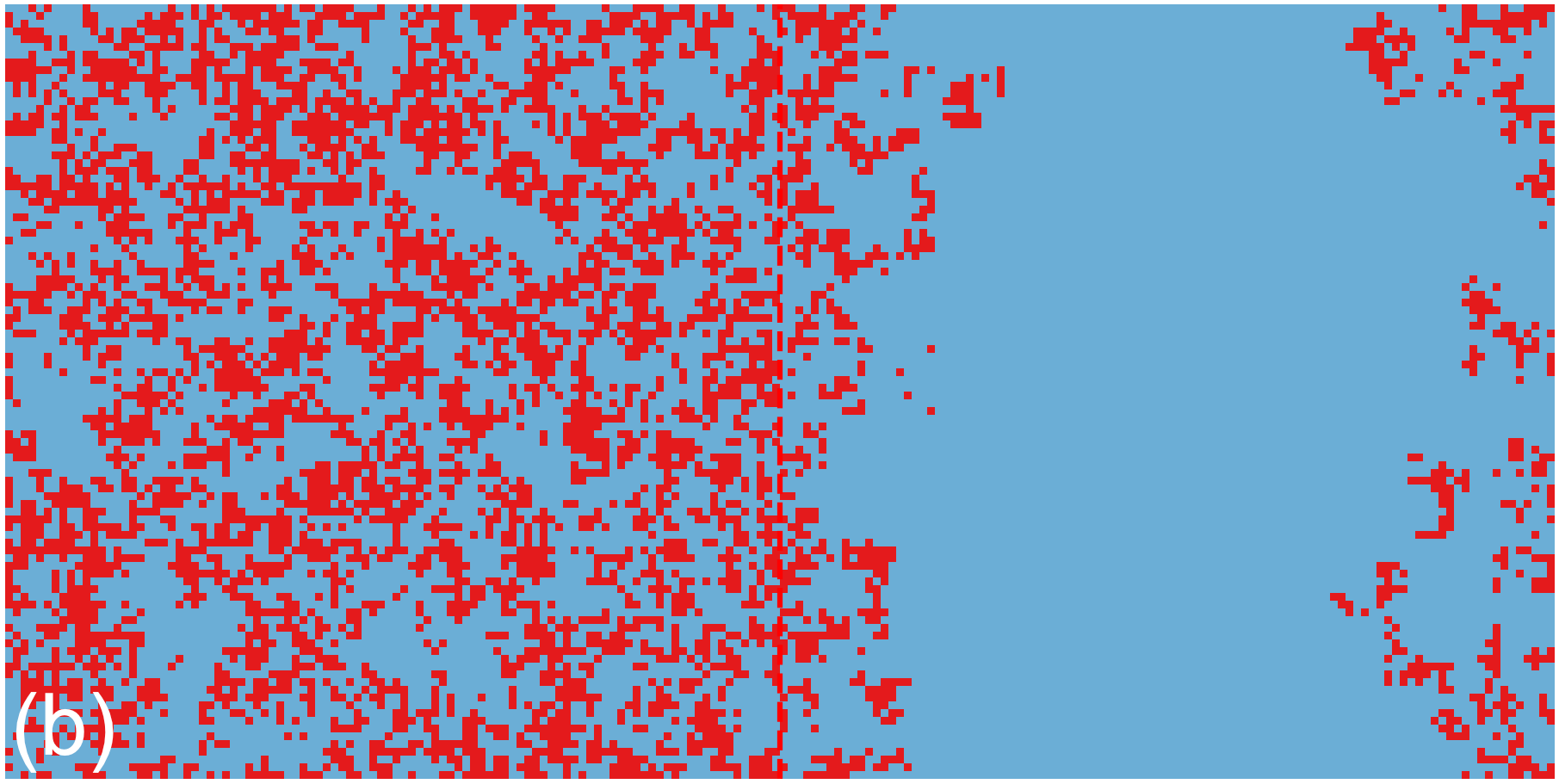} \,
\includegraphics[width=0.47\columnwidth]{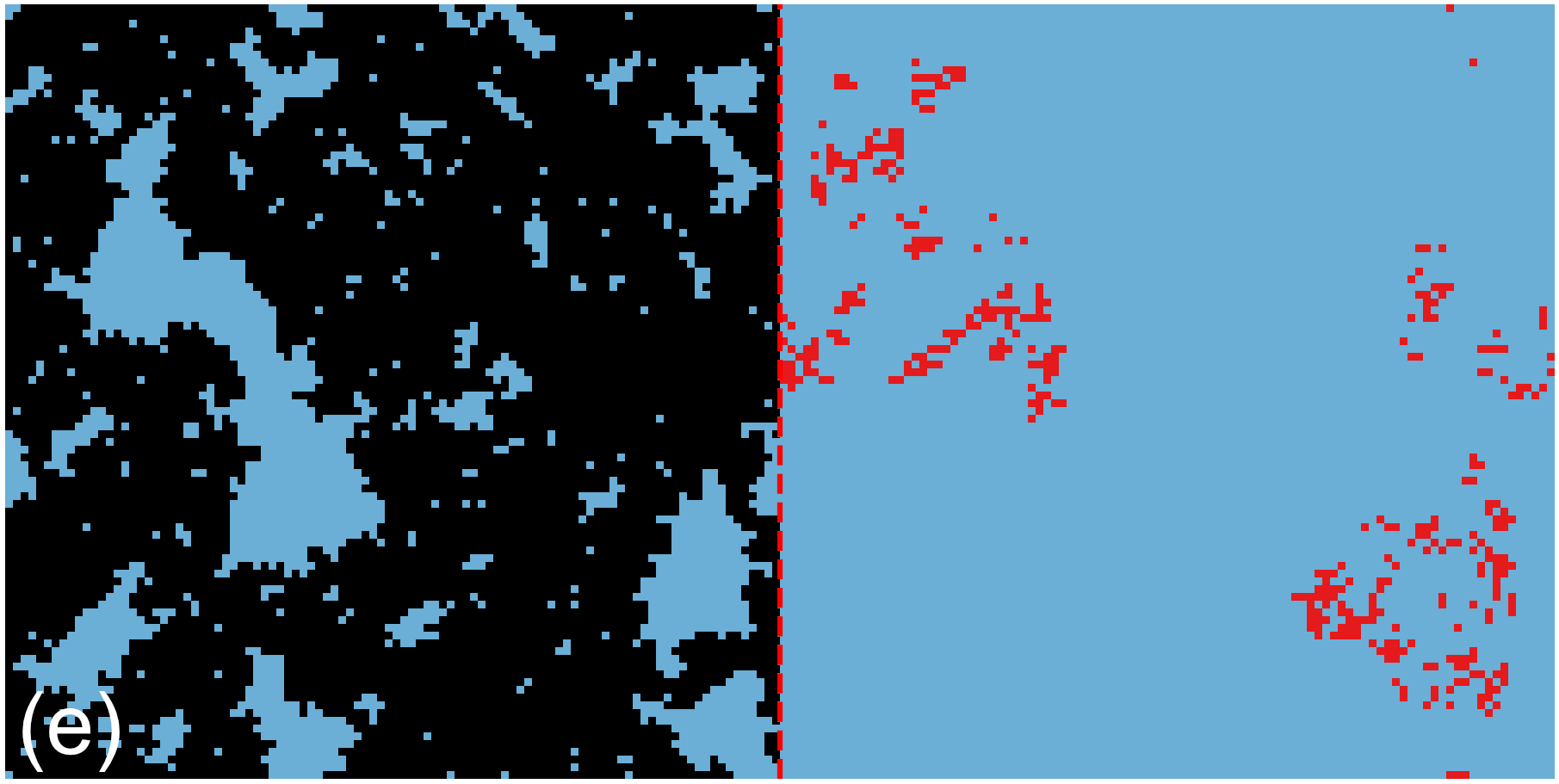} \\ \smallskip
\includegraphics[width=0.47\columnwidth]{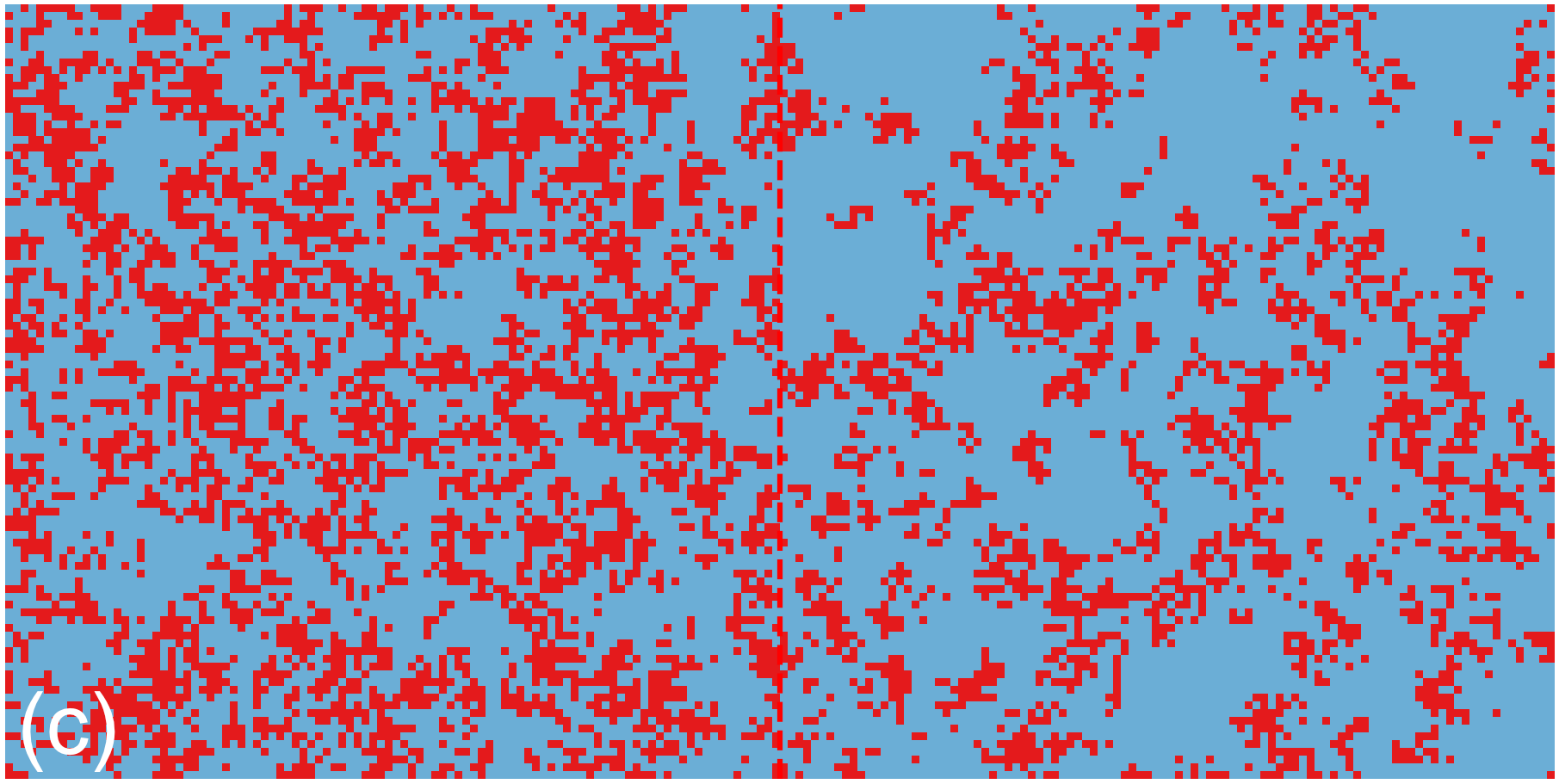} \,
\includegraphics[width=0.47\columnwidth]{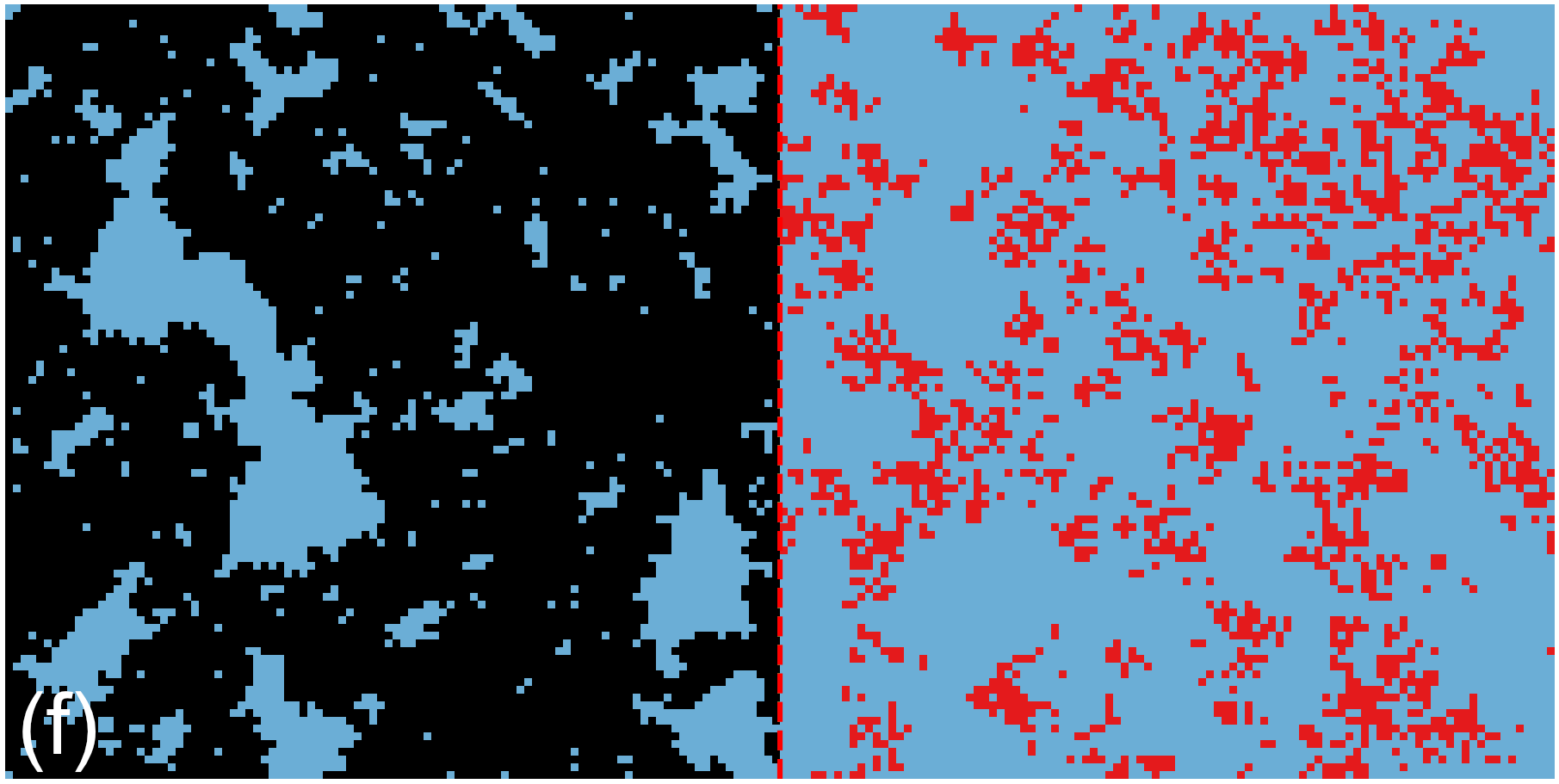} \vskip -0.1cm
\caption{Snapshots of a single Monte Carlo simulation run for two diffusively coupled SIS
	(left column) and SIR (left) / SIS (right) regions (right column) of equal size on a square 
	lattice ($L_\parallel = L_\perp = 100$, periodic boundary 	conditions, carrying capacity 
	$K = 1$), initialized ($t = 0$) with a single infectious (red) seed immersed in a 
	susceptible (blue) population with fixed recovery probability $\gamma = 0.1$.
	For the coupled SIS--SIS system (a--c), the infection probabilities are $\lambda = 0.115$
	(left region) and $\lambda = 0.109$ (right subsystem), which positions the right patch 
	very close to the epidemic threshold.
	(a) By $t = 1300$ MCS, the epidemic has spread in the left subsystem, but has gone 
	extinct on the right side.
	(b) At $t = 2900$ MCS, the spanning infection cluster has reached the active left region's 	boundaries and spreads into the inactive, susceptible domain (right).
	(c) Although the diffusive coupling was turned off at $t = 3000$ MCS, in the 
	(quasi-)stationary regime ($t = 6000$ MCS), both isolated regions sustain the 
	epidemic outbreak, with a lower infectious population density $n_I$ on the right.
	The combined SIR--SIS model (d--f) is subject to the respective infection probabilities 
	$\lambda = 0.210$ (left region) and $\lambda = 0.109$ (right) (on the left, recovered 
	individuals are indicated black).
	(d) At $t = 400$ MCS, the epidemic is spreading in the left SIR patch, but has already
	gone extinct on the right SIS side.
	(e) The snapshot at $t = 2000$ MCS depicts extended infection clusters emanating from
	the interfaces and invading into the inactive susceptible SIS region (right); the outbreak 
	has run its course in the SIR region and reached its final state with a majority of 
	recovered sites (black) interspersed by untouched susceptible (blue) patches.
	(f) At longer run times (here $t = 6000$ MCS), both effectively decoupled systems 
	reside in their respective (quasi-)stationary states, with infection events occuring 
	continuously and persistently in the right SIS patch.
	The insets respectively show the population density increases for (a) $n_I(t)$ in both 
	subsystems (averaged over the surviving runs on the more robust left side), and 
	(e) $n_I(t) + n_R(t)$ (SIR region) and $n_I(t)$ (SIS patch) \cite{Shabani26, Movies}.
\label{fig:SISsnp}} \vskip -0.6cm
\end{figure}
As above for the LV system, in Fig.~\ref{fig:SISsnp}(a--c) we explore the ensuing stochastic 
dynamics in a heterogeneous combination of two now equally-sized square diffusively 
coupled SIS regions.
The stable, active epidemic state ($\lambda = 0.115$, left side) and a vulnerable, 
extinction-prone subsystem ($\lambda = 0.109$, on the right) are initiated with all sites 
occupied by susceptibles $S$ (blue) except for a single $I$ (red) individual placed at the 
subdomains' centers.
In the run shown here, we separate and thus isolate both subsystems at $t = 3000$ MCS.
As expected, at $t = 1300$ MCS, one observes a near-threshold (fractal) spreading infection 
cluster on the left, whereas the infectious wave has terminated on the right, 
Fig.~\ref{fig:SISsnp}(a).
Once the epidemic wave reaches both interfaces to the inactive area, it incites {\em pervading 
infectious clusters} (rather than well-defined fronts) that extend deep into the excitable 
regions ($t = 2900$ MCS) and continue to do so even long after the diffusive coupling 
between the domains was shut down.
Ultimately ($t = 6000$ MCS), both SIS subsystem reach their (quasi-)steady states with 
sustained epidemic spreading, but naturally a higher ratio of infected individuals in the patch 
with larger $\lambda$ \cite{Shabani26, Movies}.

An intriguing variation is explored in Fig.~\ref{fig:SISsnp}(d--f) where we diffusively couple 
the previous extinction-prone SIS system to a domain governed by the SIR model 
\cite{Kermack27, Murray02, Keeling11} stochastic processes $S + I \to I + I$ (probability 
$\lambda$) and $I \to R$ (probability $\gamma$), where no return from ``recovered'' to 
susceptible states $R \to S$ is allowed \cite{Shabani26, Movies}.
As a consequence, only a {\em single} epidemic wave can be sustained in the SIR system,
and the infection ultimately dies out irrespective of the model parameters. 
We again combine two square SIR and SIS subsystems of the same linear size $L = 100$ 
initiated ($t = 0$) with a single active $I$ (red) site each; on the SIS side (right), we set 
$\lambda = 0.109$ and $\gamma = 0.1$.
For the SIR region, we impose $\lambda = 0.210$ and $\gamma = 0.1$, which allows the
epidemic wave to spread across the entire domain.
In this Monte Carlo run, the infection has gone extinct by $t = 400$ MCS in the SIS patch.
Subsequently, the infectious cluster spreads on the SIR patch, leaving in its wake large 
recovered areas (black) with smaller untouched susceptible (blue) enclosures.
After reaching the interfaces to the SIS subsystem, it spawns new infections in the excitable 
SIS domain.
By $t = 2000$ MCS, the SIR outbreak has ceased, terminating the {\em transient} excitations 
across the boundaries to the SIS side; still the epidemic there keeps spreading and eventually 
reaches its (quasi-)stationary state ($t = 6000$ MCS).
Similarly, the single outbreak in an SIR region where the infectious cluster reaches across the 
entire system can trigger an epidemic in a diffusively coupled SIR patch with lower infectivity 
for which a single active $I$ seed could not spawn sustained disease spreading 
\cite{Shabani26, Movies}.

% {\em Summary and outlook.} 
Illustrated by several paradigmatic models from ecology, epidemiology, and game theory, we
have demonstrated that instabilities toward stochastic extinction or fixation can be effectively 
countermanded by diffusively coupling the vulnerable subsystem to an even much smaller 
stable region.
Originating at the interfaces, influx through immigration waves effectively resets initial
conditions and suffices to maintain activity across the entire system, even when the two 
patches are eventually decoupled.
We emphasize that this boundary-driven stabilization should be a quite generic phenomenon
in finite complex systems tending towards an absorbing state.
The basic requirements for this mechanism only appear to be that (1) typical extinction
times depend sensitively on system size; (2) the source region stays stable during the time 
period when both subsystems are diffusively coupled; and (3) the absorbing region remains 
``excitable", i.e., capable of sustaining traveling invasion fronts.
Hence we expect this intriguing scenario to be broadly applicable to a wide range of natural 
phenomena. 
An experimental realization is conceivable in bacterial range expansion in a chemostatted, 
spatially inhomogeneous environment.
\smallskip

%\begin{acknowledgments}
We gratefully acknowledge helpful discussions with Erwin Frey and Michel Pleimling.
We are indebted to Shannon Serrao and Canon Zeidan for providing unpublished
preliminary data on their May--Leonard model Monte Carlo simulations.
This research was supported by the U.S. National Science Foundation, Division of Mathematical 
Sciences under Award No. NSF DMS-2128587; and sponsored by the Army Research Office 
and was accomplished under Grant Number W911NF-26-1-A323.
%\end{acknowledgments}

\end{document}